Author:
Maxwell J. D. Ramstead (maxwell.d.ramstead@gmail.com)

Affiliations:
1. Wellcome Centre for Human Neuroimaging, University College London, London, UK.
2. VERSES Labs/Spatial Web Foundation, Los Angeles, California, USA.

Word counts:
Abstract: 60 words
Main text: 1000 words
References: 1174 words
Entire text: 1446 words

Title:
The empire strikes back: Some responses to Bruineberg and colleagues.

Abstract:
In their target paper, Bruineberg and colleagues provide us with a timely opportunity to discuss the formal constructs and philosophical implications of the free-energy principle. I critically discuss their proposed distinction between Pearl blankets and Friston blankets. I then critically assess the distinction between inference with a model and inference within a model in light of instrumentalist approaches to science.

Acknowledgements:
I wish to express my gratitude to Lancelot Da Costa, Karl Friston, Conor Heins, Alex Kiefer, Dalton Sakthivadivel, and Thomas Parr for helpful discussions that were of great assistance in writing this paper.

Funding statement:
No funding to report.

Conflicts of Interest statement:
None.

Main text:

Bruineberg and colleagues provide us with a timely opportunity to discuss the core mathematical and philosophical aspects of the variational free-energy principle (FEP) and the Bayesian mechanics that follows from it. They focus on the construct of Markov blankets (MBs), which they claim has been deployed in two different—but largely conflated—ways in the literature. In their view, this conflation has led some to unduly project the epistemic virtues of one use of the construct—to formalize conditional independence in the context of inference in Bayesian networks, what they call "Pearl blankets" (PBs)—to another use; namely, to demarcate the boundaries of things that exist—what they call "Friston blankets" (FBs). Furthermore, the authors argue that these constructs are employed to pursue quite distinct research projects. These are "inference with a model," where the features of the model (here, the MBs) are assumed to be part of the scientist's model of the world, and "inference within a model," where these features are assumed to be present in the modelled system itself (i.e., the MB is cast as an actually existing boundary between the system modelled and its embedding environment).

Here, I respond critically to these claims. First, I argue that PBs and FBs are nothing more than articulations of Markov blankets in different mathematical contexts (i.e., in static statistical inference versus stochastic dynamics). Second, I claim that an instrumentalist reading of the FEP is available and informative, and that it is not given enough consideration in the account by Bruineberg and colleagues.

1. Friston blankets and Pearl blankets are just Markov blankets

The target paper claims that FBs are a novel mathematical construct that do not inherit their epistemic virtues from the use of MBs in statistical inference (i.e., from the use of PBs). Mathematically speaking, however, there is no justification for this hardline distinction.

This is because FBs and PBs *just are* MBs—in different in mathematical contexts. Generally speaking, as the authors note, MBs formalize conditional (in)dependence between variables, where the MB itself is a set of variables that renders two other sets of ('internal' and 'external') variables conditionally independent of each other. The fundamental distinction is that PBs are the kind of MB that arises in the context of static statistical inference within a Bayes network, while FBs arise when considering the interdependencies between dynamics, which is crucial, for instance, when defining the self in relation to the non-self. However, both PBs and FBs are MBs.

A same mathematical object may have different mathematical properties in different mathematical contexts. Consider an analogy. Let us list the numbers that are generated by the Peano axioms. We start by defining a first number, and we call it "0." Next, we define a successor function $S$, such that for ever number $n$, $S(n) = n + 1$. Now, when considered as objects of the *category of natural numbers*, these Peano numbers have specific properties. For instance, there exist no numbers between any numbers and its successor, and there exists no number smaller than zero. Next, consider the same sequence, now interpreted in the *category of real numbers*. Although the objects have not changed (since, after all, we are considering the same sequence), their properties are remarkably different. For instance, there are not zero but now infinitely many numbers between any number in the sequence and its successor, and there are infinitely many numbers smaller than zero.

This example illustrates that mathematical context matters in determining the properties of mathematical objects. Just as the number 1 is the same object in both categorical interpretations (albeit with different properties when considered as a natural versus as a real number), so, too, are FBs and PBs—just MBs, albeit defined in different mathematical contexts. In category-theoretic terms, we have changed the category, so the object instance has new properties, but it is still the same object.

2. The Bayesian mechanics is physics, not metaphysics

I believe that the core issue with the target paper is that it treats the FEP as if it were a metaphysical statement, when in reality, it is better understood as a new chapter of physics—a Bayesian mechanics—and is compatible with instrumentalism about scientific theories (Friston, Heins, Ueltzhoffer, Da Costa, & Parr, 2021). Although the authors discuss this possibility in passing, they downplay its significance.

There is a longstanding tradition in the philosophy of physics that is *instrumentalist* about physical theories (Giere, 1999, 2010; Van Fraassen, 1980). On the instrumentalist view, all scientific models, such as the ones used in physics, are literally false and play the role of useful fictions that help us to understand the world.

Now, there is nothing about the FEP that commits us to realism about scientific models. In fact, we have argued that precisely the opposite is the case; see (Ramstead, Friston, & Hipólito, 2020).

In section 3.2. of their paper, the authors provide a great outline of the modelling strategy at play in the computational neuroscience literature: what they call "models of models." We agree with this way of putting things but would suggest to extend the logic at play to FBs. They point out that in computational neuroscience, there are two 'levels' of modelling at play, which are explored in (Ramstead et al., 2020). The first level is that of the scientist that is modelling some target phenomenon, e.g., constructing scientific models of living systems. The second level is that of the system being modelled. In computational neuroscience, scientists are constructing *scientific models of the inferential models and processes* that are assumed to be used by cognitive systems themselves. Crucially, this 'second level' of modelling is just a special case of the first level—it just happens to be that the physical system being modelled, is *modelled as if* it were inferring the causes of its sensory states. As Alex Kiefer put it (personal communication), according to the FEP, the best scientific model of the organism is a statistical model of its world.

From an instrumentalist perspective, there is no robust philosophical difference between inference with a model and inference within a model. There is no reason why we cannot use MBs as a modelling tool, to carve out the boundaries of systems, when these are modelled as random dynamical systems (i.e., as sets of random variables over paths—i.e., stochastic processes—with dependence relations).

The authors appeal to the notion of a *formal ontology* that was introduced by (Ramstead, Constant, Badcock, & Friston, 2019), but misapprehend it. The formal ontology that flows from the Bayesian mechanics is not an *a priori* attempt to find or draw boundaries in nature. Rather, it is an attempt to construct *scientific models of these boundaries*, in an *instrumentalist* fashion. The formal ontology only entails that we create empirically evaluable, formal models of organism boundaries. Interestingly, this model of boundaries is itself evaluable via Bayesian model evidence, leading to a nice, reflexive aspect to the framework: our scientific model of the boundaries of living systems that can be improved iteratively via free energy or prediction error minimization.